\documentclass[twocolumn,amsmath,amsfonts,amssymb,showpacs]{revtex4} 

\usepackage{graphicx}
\def\beq{\begin{equation}}
\def\eeq{\end{equation}}
\def\bea{\begin{eqnarray}}
\def\eea{\end{eqnarray}}

\newcommand{\ket}[1]{|#1\rangle}
\newcommand{\bra}[1]{\langle #1|}
\begin{document}

\title{Entanglement of arbitrary spin modes in an expanding universe}

\author{Zahra Ebadi}
\affiliation{Department of Physics, Isfahan University of Technology, Isfahan 84156-83111, Iran}
\author{ Hossein Mehri-Dehnavi}
\affiliation{Department of Physics, Babol University of Technology, Babol, 47148-71167, Iran.}
\author{Behrouz Mirza}
\affiliation{Department of Physics, Isfahan University of Technology, Isfahan 84156-83111, Iran}
\author{Hosein Mohammadzadeh}
\affiliation{Department of Physics, University of Mohaghegh Ardabili,  P.O. Box 179, Ardabil, Iran.}
\email{mohammadzadeh@uma.ac.ir}
\author{Robabeh Rahimi}
\affiliation{Institute for Quantum Computing, University of Waterloo, Waterloo, ON, N2L 3G1,
  Canada.}

\pacs{05.20.-y, 67.10.Fj}

\begin{abstract}
Pair particle creation is a well-known effect on the domain of field
theory in curved space-time. 
It is  shown that the entanglement generations for spin-0 and
spin-1/2 modes are different in Friedmann-Robertson-Walker (FRW)
space-time. We consider the spin-1 particles in FRW space-time using
Duffin-Kemmer-Petiao (DKP) equation and obtain a measure of the
generated entanglement. Also, we consider the spin-3/2 particles. We
argue that the absolute value of the spin does not play any role in
entanglement generation and the differences are due to the bosonic
or fermionic properties.
\end{abstract}
\maketitle

\section{Introduction}
Because of the importance and the fundamental role of entanglement
in quantum information processing  and quantum computing, a lot of
researches have been done on entanglement generation, variation and
degradation in various domains. In non-relativistic limits,
entanglement has been extensively studied. Recently, relativistic
quantum information processing has attracted a lot of interests
\cite{peres,alsing1,shi,alsing2}.  The world is fundamentally
relativistic, therefore, understanding entanglement in space-time is
ultimately important. It is realized that relativity plays a
significant role in quantum entanglement and related quantum
protocols, such as quantum teleportation. This point is justified by
quantum optics, which is well established on the basis of not only
quantum theory but also special relativity in nature. Most of
EPR-type experiments have been performed by photon pairs. In
addition, experiments of quantum teleportation have been extensively
carried out by photons \citep{He,Aspect,Weihs,Bouwmeester,Furusawa}.

There are some qualitative differences between non-relativistic
entanglement and relativistic one. It has been shown that
entanglement is an observer dependent property in non-inertial
frames. I. Fuentes {\it et. al.} have showed that the entanglement
of both bosonic and fermionic modes degrade by acceleration
\cite{Fuentes1,Fuentes2}. In addition, quantum discord and
entanglement of pseudo-entangled spinor modes in non-inertial frames
have been studied \cite{mehri}.

Recently, scalar and spinor field modes have been investigated in an
expanding universe. It has been shown that in both cases, the
expanding universe generates entanglement between field modes \cite{footnote}.

 A separable vacuum state in distant past time  appears
as an entangled state in far future time because of the expanding
universe. However, there are differences between the entanglement of
scalar boson fields and spinor fermion fields
\cite{Fuentes3,Fuentes4}.

It has been shown that the entanglement of massive boson modes is a
monotonically decreasing function  with respect to the momentum of
the modes while for the fermion modes, there is an optimum point for
the momentum of the modes. For the spin-1/2 case, there is no
entanglement between the zero momentum modes while for the spin-0
case the maximum entanglement is related to the zero momentum modes.
As a common behavior between the spin-0 and spin-1/2 modes, it has
been shown that there is no entanglement for massless bosons or
fermions \cite{Fuentes4}.

To understand the origin of the differences between the generated
entanglement in an expanding universe for spin-0 and spin-1/2 modes,
we  investigate the higher spin modes. Firstly, we consider the
spin-1 modes using Friedmann-Robertson-Walker (DKP) equation in
Duffin-Kemmer-Petiao (DKP) space-time and work out a measure for the
generated entanglement. The same procedure will be applied on
spin-3/2 modes using Rarita-Schwinger equation.

This paper is organized as follows: In section II, we consider the
DKP equation in FRW  expanding universe specially in two dimensions.
In section III, we calculate the entanglement entropies of spin-1
and spin-0 particles and compare them to each other. Also, we
investigate the variation of entanglement with respect to the
parameters of expansion, and the momentum of any the mode. In
section IV, we
 consider the spin-3/2 particles using Rarita-Schwinger equation
and work out the generated entanglement and compare it with the
entanglement of spin-1/2 modes. Conclusions are presented in section
V.

\section{DKP equation in Friedmann-Robertson-Walker space-time}\label{2}
It is well-known that Klein-Gordon and Dirac equation describe
particles with spin-0 and spin-1/2 in flat Minkowski space-time,
respectively. Scalar and spinor fields have been considered in more
details in curved space-time. There are various ways of formulating
a relativistic wave equation describing the dynamical states of a
massive vector boson, such as Proca equation, Duffin-Kemmer-Petiau
equation and Weinberg-Shay-Good equation
\cite{kemmer,weinberg,shay,farhad}. We employ DKP equation for
considering vector bosons in curved space-time. Before starting the
study of the DKP in curved space-time, we notice that it is similar
to the Dirac equation in Minkowski space-time as follows:
 \bea
 (i\beta^{\mu}\partial_{\mu}-m)\Psi=0,
 \eea
where, the $\beta^{\mu}$-matrices are generalization of the Dirac
gamma matrices, satisfying an algebra ring, which for spin-1 is
 \bea
 \beta^{\lambda}\beta^{\mu}\beta^{\nu}+\beta^{\nu}\beta^{\mu}\beta^{\lambda}=\eta^{\lambda\mu}\beta^{\nu}+\eta^{\mu\nu}\beta^{\lambda}.
 \eea
Generally, $\beta^{\mu}$'s represent four $16\times16$ reducible
matrices, which decompose into three separate representations, a one
dimensional trivial, a five dimensional spin-0 and a ten dimensional
spin-1 representations \cite{mirza,swansson,falek,greiner}. For the
the 10-dimensional spin-1 representation,  $\beta^{\mu}$ matrices
given by
 \bea
 \beta^{0}&=&\left(\begin{array}{cccc}
                   0 & \textbf{0}_{1\times3} & \textbf{0}_{1\times3} & \textbf{0}_{1\times3} \\
                   \textbf{0}_{3\times1} & \textbf{0}_{3\times3} & \textbf{1}_{3\times3} & \textbf{0}_{3\times3} \\
                   \textbf{0}_{3\times1} & \textbf{1}_{3\times3} & \textbf{0}_{3\times3} & \textbf{0}_{3\times3} \\
                   \textbf{0}_{3\times1} & \textbf{0}_{3\times3} & \textbf{0}_{3\times3} & \textbf{0}_{3\times3}
                 \end{array}
                 \right),\\
 \beta^{i}&=&\left(\begin{array}{cccc}
                   0 & \textbf{0}_{1\times3} & {{\textbf{K}^{j}}} & \textbf{0}_{1\times3}  \\
                   \textbf{0}_{3\times1} & \textbf{0}_{3\times3} & \textbf{0}_{3\times3} & -i{\textbf{S}}^{j} \\
                   -{{\textbf{K}^{j}}}^{\dagger} & \textbf{0}_{3\times3} & \textbf{0}_{3\times3} & \textbf{0}_{3\times3} \\
                    \textbf{0}_{3\times1} &  -i{\textbf{S}}^{j} & \textbf{0}_{3\times3} & \textbf{0}_{3\times3}
                 \end{array}
                 \right),
 \eea
where, $\textbf{S}^{i}$'s are the standard $(3\times3)$ spin-1
matrices and $\textbf{K}^{i}$'s denote  $(1\times3)$ matrices with
elements $\textbf{K}^{i}_{j}=\delta^{i}_{j}$.

In curved space-time, we can use the tetrad formalism to obtain the generalized DKP equation
 \bea
 \left(i{\tilde{{\beta}}}^{\mu}(\partial_{\mu}+\frac{1}{2}\omega_{\mu ab}\mathcal{S}^{ab})-m\right)\Psi=0,
 \eea
where, $\mathcal{S}^{ab}=[\beta^{a},\beta^{b}]$ and
${\tilde{{\beta}}}^{\mu}$'s are the Kemmer matrices in curved
space-time and they are related to Minkowski space-time
${\tilde{{\beta}}}^{\mu}=e^{\mu}_{a}\beta^{a}$ with the following
tetrad relation
 \bea
 {e^{\mu}}_{a}{e^{\nu}}_{b}\eta^{ab}=g^{\mu\nu},~~~  {e^{\mu}}_{a}e_{b\mu}=\eta_{ab},~~~{e^{\mu}}_{a}{e_{\mu}}^{b}=\delta^{b}_{a}.
 \eea
Also, the spin connections $\omega_{\mu ab}$ are given by
 \bea
 \omega_{\mu ab}=e_{a l}{e^{j}}_{b}\Gamma^{l}_{j\mu}-{e^{j}}_{b}\partial_{\mu}e_{a j},
 \eea
where, $\Gamma^{l}_{j\mu}$'s are the affine connections, which are
obtained by the space-time metric elements. Specifically, we
consider a two dimensional FRW expanding space-time with line
element
 \bea
 ds^{2}=C^{2}(\eta)(d\eta^{2}-dx^{2}),\label{metric}
 \eea
where, $\eta$ is the conformal time, and the conformal scale factor,
$C$, is given by
 \bea
 C(\eta)=\left(1+\epsilon(1+\tanh\rho\eta)\right)^{1/2},\label{RW}
 \eea
with positive real parameters $\epsilon$ and $\rho$, controlling the
total volume and rapidity of the expansion. Primarily, the
entanglement between the modes of a quantum field in a curved
space-time can be investigated in special states where the
space-time has at least two asymptotically flat regions. According
to Eq. (\ref{RW}), in the distant past and far future, the
space-time becomes Minkowskian, since $C(\eta)$ tends to
$1+2\epsilon$ for $\eta\rightarrow +\infty$ and it tends to $1$ for
$\eta\rightarrow-\infty$. In the intermediate region, the concept of
the particle breaks down.

For simplicity, we restrict ourselves to solve the
($1+1$)-dimensional DKP equation in FRW space-time. The DKP equation
can be obtained in the following form \cite{falek1}
 \bea
 \left[\beta^{0}\partial_{\eta}+ik\beta^{1}-\frac{\dot{C}}{C}{(\beta^{1})}^{2}\beta^{0}+iCm\right]\tilde{\Psi}=0,
 \eea
where, $\Psi(\eta,x)=e^{i\vec{k}.\vec{x}}\tilde{\Psi}(\eta)$ and
$\tilde{\Psi}(\eta)$ has ten components as follows
 \bea
 \tilde{\Psi}(\eta)=\left(
                      \begin{array}{c}
                        \varphi \\
                        P \\
                        Q \\
                        R \\
                      \end{array}
                    \right),
 \eea
where, $P, Q$ and $R$ are $3\times1$ vectors and $\varphi$ denotes a
scalar. The Kemmer matrices representation for a suitable
arrangement of the components of these vectors and $\varphi$ leads
to the following equations
 \bea
 &&iCm\Phi=-(\partial_{\eta}+\frac{\dot{C}}{C})\mathcal{X}+ik\Theta,\nonumber\\
 &&iCm \mathcal{X}=-\partial_{\eta}\Phi,\\
 &&iCm\Theta=-ik\Phi.\nonumber
 \eea
Where
 \bea
 \Phi=\left(\begin{array}{c}
        P_{2}   \\
        P_{3}   \\
        Q_{1}
      \end{array}\right),~~~~\mathcal{X}=\left(\begin{array}{c}
                             Q_{2} \\
                             Q_{3} \\
                             P_{1}
                           \end{array}\right),~~~~
                           \Theta=\left(
                                    \begin{array}{c}
                                      R_{3} \\
                                      -R_{2} \\
                                      \varphi \\
                                    \end{array}
                                  \right).
 \eea
The third component of $R$ is vanished. We can obtain an independent
equation for $\Phi(\eta)$ as follows
 \bea
 \left(\frac{d^{2}}{d\eta^{2}}+k^{2}+C^{2}m^{2}\right)\Phi(\eta)=0,\label{phi}
 \eea
 and $\mathcal{X}(\eta)$ and $\Theta(\eta)$ satisfy the following equations
 \bea
 \mathcal{X}=\frac{i}{Cm}\partial_{\eta}\Phi,~~~~~~\Theta=-\frac{k}{Cm}\Phi.\label{KT}
 \eea
According to the FRW matrices and using Eq. (\ref{phi}) we obtain
the following two different solutions, which  are analytic in $in$
region, $\Phi_{in}$, for distant past,  and  $out$ region,
$\Phi_{out}$, for far future.
 \begin{widetext}
 \bea
 &&\Phi_{in}(\eta)=\left(\frac{1}{2}(1+\tanh\rho\eta)\right)^{\frac{i}{2\rho}\omega_{in}}
 \left(\frac{1}{2}(1-\tanh\rho\eta)\right)^{\frac{i}{2\rho}\omega_{out}} {_{2}F_{1}}\left(\alpha,\beta,\gamma,\frac{1}{2}(1+\tanh\rho\eta)\right)\textbf{V},\\
 &&\Phi_{out}(\eta)=\left(\frac{1}{2}(1+\tanh\rho\eta)\right)^{\frac{i}{2\rho}\omega_{in}}
 \left(\frac{1}{2}(1-\tanh\rho\eta)\right)^{\frac{i}{2\rho}\omega_{out}} {_{2}F_{1}}\left(\alpha,\beta,1+\alpha+\beta-\gamma,\frac{1}{2}(1-\tanh\rho\eta)\right)\textbf{V},
 \eea
 \end{widetext}
where
 \bea
 && \alpha=1+\beta,~~~~\beta=\frac{i}{2\rho}(\omega_{in}+\omega_{out}),~~~~\gamma=1+\frac{i\omega_{in}}{\rho}\nonumber\\
 &&\omega_{in}^2={k^2+m^2(1+2
\epsilon)},~~~~\omega_{out}^2={k^2+m^2},
 \eea
and $\textbf{V}$ is a constant vector of dimension $3\times1$.
$\mathcal{X}(\eta)$ and $\Theta(\eta)$ can be evaluated easily from
(\ref{KT}). $ _2F_1(a,b,c,d)$'s are  hypergeometric functions. Using
these functions properties, the above solutions will be related to
each other by well-known Bogoliubov transformation technique as
 \bea
 \Phi_{in}=\alpha_{k}\Phi_{out}+\beta_{k}\Phi_{out}^{\ast},
 \eea
where, $\alpha_{k}$ and $\beta_{k}$ are Bogoliubov coefficients.
Since $\mathcal{X}(\eta)$ and $\Theta(\eta)$ are given by
$\Phi(\eta)$, one can relate the $in$ and $out$ modes of the
wavefunction by a similar equation as follows

  \bea
 \Psi_{in}=\alpha_{k}\Psi_{out}+\beta_{k}\Psi_{out}^{\ast}.
 \eea
Using the properties of hypergeometric functions \cite{handbook}, we evaluate the
Bogoliubov coefficients
 \bea
 \alpha_{k}=\frac{\Gamma\left(1+\frac{i\omega_{in}}{\rho}\right)\Gamma\left(-\frac{i\omega_{out}}{\rho}\right)}
 {\Gamma\left(1+\frac{i(\omega_{in}-\omega_{out})}{2\rho}\right)\Gamma\left(\frac{i(\omega_{in}-\omega_{out})}{2\rho}\right)}, \label{alpha}\\
 \beta_{k}=\frac{\Gamma\left(1+\frac{i\omega_{in}}{\rho}\right)\Gamma\left(\frac{i\omega_{out}}{\rho}\right)}
 {\Gamma\left(1+\frac{i(\omega_{in}+\omega_{out})}{2\rho}\right)\Gamma\left(\frac{i(\omega_{in}+\omega_{out})}{2\rho}\right)}.\label{beta}
 \eea
Now, we can  work out a relation between the annihilation and
creation operators in $in$ and $out$ regions as follows
    \bea
    a^{in}_{k}=\alpha^{*}_{k}a^{out}_{k}+\beta^{*}_{k}{b^{\dagger}_{k}}^{out},\label{bugulibov}
    \eea
where, $a$ and $a^{\dagger}$ ($b$ and $b^{\dagger}$) are the
annihilation and creation operators of particles (anti-particles)
and satisfy the following well-known relations
     \bea
    [{a^{\dagger}_{k}}^{in(out)},a_{k^{\prime}}^{in(out)}]=\delta(k-k^{\prime}),\nonumber\\
    {[{b^{\dagger}_{k}}^{in(out)},b_{k^{\prime}}^{in(out)}]}=\delta(k-k^{\prime}).
    \eea

The five dimensional representations of Kemmer matrices are given by
 \bea
 \beta^{0}=\left(\begin{array}{cc}
                   {\sigma}_{x} & \textbf{0}_{2\times3}  \\
                   \textbf{0}_{3\times2} & \textbf{0}_{3\times3}
                 \end{array}
                 \right), ~~~
 \beta^{i}=\left(\begin{array}{cc}
                   \textbf{0}_{2\times2} & \mathbf{\rho^{i}}_{2\times3} \\
                   {\mathbf{\rho^{i}}}^{\dagger}_{3\times2} & \textbf{0}_{3\times3}
                 \end{array}
                 \right),
 \eea
where, $\sigma_{x}$ is the Pauli matrix and
$\mathbf{\rho^{i}}_{jk}=-\delta_{1j}\delta_{ik}$. In this case,
$\tilde{\Psi}(\eta)$ has five ${\tilde{\Psi}}_{i}$ components, where
$i=1,\cdots,5$. Therefore, DKP equation reduces to the following
equations
 \bea
 &&i C m{\tilde{\Psi}}_{1}=-(\partial_{\eta}-\frac{\dot{C}}{C}){\tilde{\Psi}}_{2}+i k {\tilde{\Psi}}_{3},\nonumber\\
 &&i C m{\tilde{\Psi}}_{2}=-\partial_{\eta}{\tilde{\Psi}}_{1},\\
 &&i C m{\tilde{\Psi}}_{3}=-i k{\tilde{\Psi}}_{1},\nonumber
 \eea
and the components ${\tilde{\Psi}}_{4}={\tilde{\Psi}}_{5}=0$.
$\tilde{\Psi}_{1}$ satisfies Eq. (\ref{phi}), and $\tilde{\Psi}_{2}$
and $\tilde{\Psi}_{3}$ satisfy the same Eqs. (\ref{KT}).
%
Therefore, the solutions of the above equations with respect to the
$in$ and $out$ modes are the same as for spin-1 case and they are
related to each other by the same Bogoliubov coefficients. This is a
significant point achieved in the rest of the paper.

\section{Entanglement Generation duo to  expansion}
We consider the vacuum state in $in$ region  as a separable state
\cite{Fuentes3}
     \bea
     \ket{0}^{in}=\prod_{k\in \mathbb{R^{+}}}
    \ket{0_{k}}^{in}\ket{0_{-k}}^{in}.
    \eea
For simplicity, we focus on a specific momentum, $k$, mode as
follows
    \bea
    \ket{0}^{in}_{k}=\ket{0_{k}}^{in}\ket{0_{-k}}^{in}.
    \eea
In other words, we disregard all other modes of vacuum by tracing
out the total density matrix over them. Using Eq. (\ref{bugulibov}),
we write the above bipartite separable state as a linear combination
of the excited states in mode $k$ by an observer in $out$ region
    \bea
    \ket{0_{k}}^{in}\ket{0_{-k}}^{in}=\sum_{n=0}^{\infty}A_{n}\ket{n_{k}}^{out}\ket{n_{-k}}^{out}.\label{in-out}
    \eea
Obviously, the last relation is the Schmidt decomposition of a pure
state of a bipartite system. The Schmidt coefficient is obtained by
the normalization condition and by applying the annihilation
operator on the state
    \bea
    a^{in}_{k}\ket{0}^{in}_{k}=(\alpha^{*}_{k}a^{out}_{k}-\beta^{*}_{k}{b^{\dagger}_{k}}^{out})\sum_{n=0}^{\infty}A_{n}\ket{n_{k}}^{out}\ket{n_{-k}}^{out}=0.\nonumber\\
    \eea
Therefore, one finds the following relation for the coefficients
    \bea
A_{n}=\left(\frac{\beta^{*}_{k}}{\alpha^{*}_{k}}\right)^{n}A_{0},
~~~~~A_{0}=\sqrt{1-\left| \frac{\beta_{k}}{\alpha_{k}} \right|^2}.
    \eea
Thus, a vacuum state in $in$ region corresponds to a state with
particle excitations in $out$ region. The particle creation via the
expansion is a well-known concept \cite{Davies}. In the following,
we concentrate on the entanglement generation related to this
concept.

We construct the density matrix in $in$ region by using Eq.
(\ref{in-out}) in $out$ region
    \bea
    \rho_{k,-k}=\ket{0_{k}}^{in}\ket{0_{-k}}^{in}\bra{0_{k}}^{in}\bra{0_{-k}}^{in}.
    \eea
Because the Schmidt coefficients in Eq.  (\ref{in-out}) are
non-zero, the $in$ vacuum is entangled from the point of view of an
observer in  $out$ region.

We use an appropriate measure of  entanglement, namely  the von
Neumann entropy  defined as follows
    \bea
    S(\rho_{k})=-{\rm tr}(\rho_{k}\log_{2}(\rho_{k})),
    \label{entropy}
    \eea
where, $\rho_{k}$ is the reduced density matrix of the particles'
subsystem
 \bea
 \rho_{k}={\rm tr}_{-k}(\rho_{k,-k}).\label{reducedmatrix}
 \eea
 The von Neumann entropy for spin-1 modes  is given by
    \bea
    S=-\sum_{n=0}^{\infty}|A_{n}|^{2}\log_{2}|A_{n}|^{2}
    =\log_{2}\frac{x^{\frac{x}{x-1}}}{1-x}\label{scalarentropy},
    \eea
where, $x=|\frac{\beta_{k}}{\alpha_{k}}|^{2}$ and $\alpha_k$ and
$\beta_k$ are  given by Eqs. (\ref{alpha}) and (\ref{beta}),
respectively. 

\begin{figure}[t]
    \center
    \includegraphics[width=0.95\columnwidth]{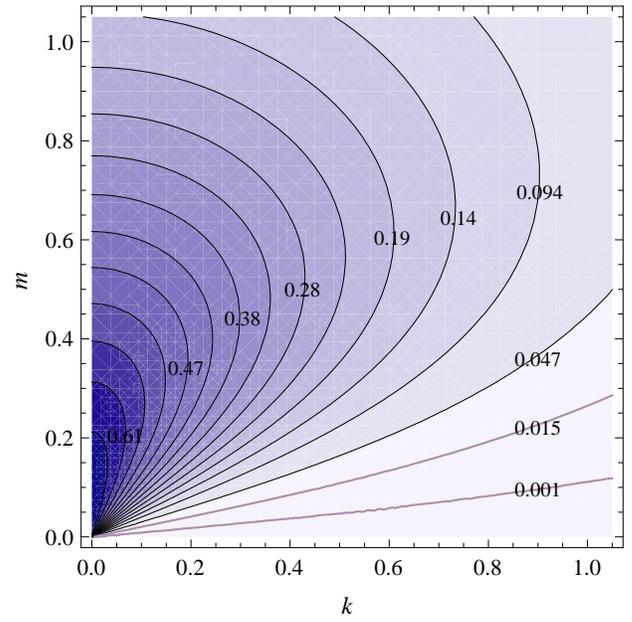}\\
    \caption{(Color online) Contour plot of the von Neumann entropy for spin-1 modes with respect to the mass and momentum for the specified expanding
    universe $\epsilon=2$ and $\rho=2$.}\label{fig1}
   \end{figure}

Fig. (\ref{fig1}) shows the contour plot of the von Neumann entropy
 with respect to the mass
and the momentum of spin-1 particles for the specified expanding
universe (fixed values of total volume and rapidity of expansion).
The generated entanglement is a decreasing function with respect to
the momentum. Also,  there is no entanglement between massless
spin-1 particles. For each value of momentum, there exists a
specific value of mass, $m_{\rm max}$, at which the entanglement is
maximum. $m_{\rm max}$ is an increasing function with respect to the
momentum.

\begin{figure}[t]
    \center
    \includegraphics[width=0.8\columnwidth]{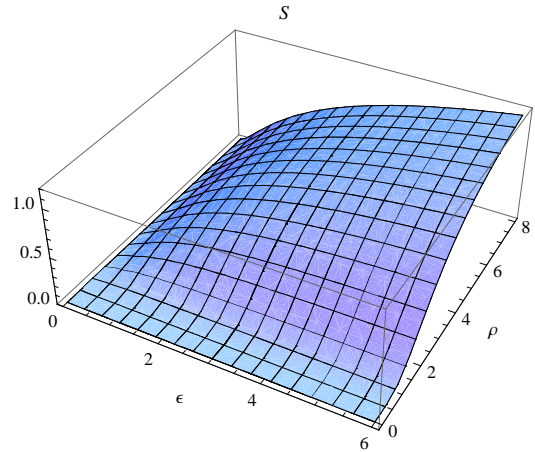}\\
    \caption{(Color online) The von Neumann entropy for spin-1 modes with respect to the total volume and rapidity of
    expansion for  $m=1$ and  $k=0.1$.}\label{fig2}
   \end{figure}

It is obvious from  Fig. (\ref{fig2}) that for the specified mass
and momentum of spin-1 particles, the generated entanglement is an
increasing function with respect to the total volume, $\epsilon$,
 and the rapidity of expansion, $\rho$. There is no entanglement
generation
  for $\epsilon=0$ and $\rho=0$, which correspond to a flat Mikowskian space-time.

\begin{figure}[t]
    \center
    \includegraphics[width=0.9\columnwidth]{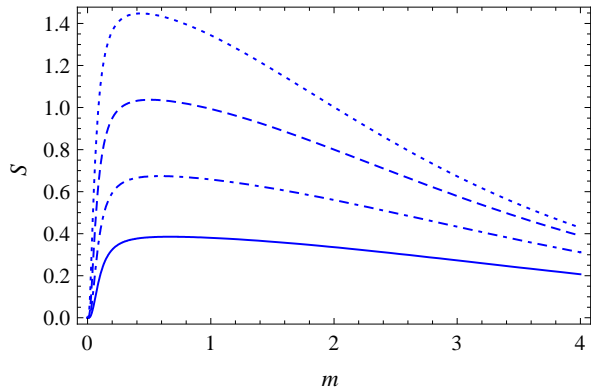}\\
    \caption{(Color online) The von Neumann entropy, $S$, for spin-1 modes with respect to mass, $m$,  for  fixed values of momentum, $k=0.1$, and  rapidity, $\rho=10$.
    The different values of total volume are $\epsilon=1$ (solid), $\epsilon=2$ (dash dotted), $\epsilon=4$ (dashed), and $\epsilon=8$ (dotted)  curves.}\label{fig3}
   \end{figure}

In addition, one can observe from  Fig. (\ref{fig3}) that there is a
certain value of mass, $m_{\rm max}$, wherein the generated
entanglement is maximum. Naturally, the entanglement generation is
larger for an expanding universe with a larger  total volume
expansion. In a similar behavior, $m_{\rm max}$  tends to a larger
value for larger total volume expansion.

The above arguments and results were represented for spin-1
particles. Since the von Neumann entropy is a function with respect
to the Bogoliubov  coefficients and we argue in section \ref{2} that
these coefficients are the same for spin-1 and spin-0 particles, the
general behavior of   entanglement generation for spin-0 particles
will be the same as spin-1 particles.

\section{Spin-3/2 particles and entanglement generation}

The Rarita-Schwinger equation characterizes the dynamics of massive
spin-3/2 particles in flat Minkowski space-time \cite{rarita}. In
supergravity models, the superpartner of graviton field is described
by spin-3/2 particles. The complicated form of the Rarita-Schwinger
equation makes it inexplicable to obtain explicit results even in a
simple background. It has been shown that when one considers
helicity $\pm 3/2$ states propagating in arbitrary homogeneous and
isotropic scalar or gravitational backgrounds, the equations can be
reduced to a Dirac-like equation in flat Minkowski space-time
\cite{velo,maroto} as follows
     \bea
(i\not{\!\partial}-m)\psi_{\mu}=0, \eea with two constraints \bea
&&\gamma^{\mu}\psi_{\mu}=0,\nonumber\\
&&\partial^{\mu}\psi_{\mu}=0.
    \eea
By replacing ordinary derivatives by covariant ones, we obtain the
equation for FRW metrics as follows \cite{maroto}
    \bea
&&(i\not{\!\! D}-m)\psi_{\mu}=0,\nonumber\\
&&\gamma^{\mu}\psi_{\mu}=0,\\
&&D^{\mu}\psi_{\mu}=0.\nonumber
    \eea
Where,
$D_{\rho}\psi_{\sigma}=(\partial_{\rho}+\frac{i}{2}\Omega^{ab}_{\rho}\Sigma_{ab})\psi_{\sigma}$
with $\Omega^{ab}_{\rho}$ the spin connection coefficients and
$\Sigma_{ab}=\frac{i}{4}[\gamma_{a},\gamma_{b}]$ and
$[D_{\mu},D_{\nu}]=-\frac{i}{2}{R^{ab}}_{\mu\rho}\Sigma_{ab}$.
According to  Eq. (\ref{metric}) for FRW expanding universe, we
obtain the spin-connections, Riemann tensor and finally the
following equation
     \bea
    \left(iC^{-1}\gamma^{\mu}\partial_{\mu}-m+i\frac{3}{2}\frac{\dot{C}}{C^{2}}\gamma^{0}\right)\psi_{\mu}=0.
    \eea
By rewriting $\psi_{\mu}$ as a multiplication of spatial part,
$\exp(i\vec{k}.\vec{x})$ and time dependent function,
$\kappa(\eta)$, also an appropriate function of $C(\eta)$, we obtain
the following equation for time dependent part \cite{maroto}
    \bea
    \left(\frac{d^{2}}{d\eta^{2}}+k^{2}-im\dot{C}+m^{2}C^{2}\right)\kappa(\eta)=0.\label{spin3/2}
    \eea
Comparison between  Eq. (\ref{spin3/2}) and the same evaluation for
spin-1/2 particles in curved space-time \cite{Duncan} shows that two
distant past and far future asymptotic solutions are related by the
same Bogoliubov coefficients. It is straightforward to show that the
von Neumann entropy is given by
    \bea
    S=-\sum_{n=0}^{1}|A_{n}|^{2}\log_{2}|A_{n}|^{2}
    =\log_{2}\frac{1+x}{x^{\frac{x}{1+x}}}\label{spinorentropy},
    \eea
where, $x=|\frac{\beta_{k}}{\alpha_{k}}|^{2}$,  $\alpha_k$ and
$\beta_k$ are Bogoliubov  coefficients. Since von Neumann entropy is
directly related to these coefficients, we can argue that the
generated entanglement of spin-3/2 particles due to the expanding
universe will have the same properties of spin-1/2 case.

\section{Conclusion}
DKP equation is employed for considering the spin-1 particles in the
FRW expanding universe. It is showed that there is two asymptotic
distant past and far future times and one can connect these
solutions to each other by the well-known Bogoliubov technique. The
separable vacuum state of distant past  time can be appeared as an
entangled particle-antiparticle state in far future time. In order
to measure the entanglement of the generated pure state,  the von
Neumann entropy is used. The general behavior of the generated
entanglement is summarized with respect to the characteristics of an
expanding universe, the mass and the momentum of any modes.

It is also shown that the general behavior of the generated
entanglement of spin-0 particles is the same as spin-1 particles
because of the same Bogoliubov coefficients which relate the
asymptotic solutions to each other.

We considered spin-3/2 particles in FRW space-time using the
Rarita-Schwinger equation. Comparing the equation which describes
the time dependent part of the field with the spin-1/2, one shows
that Bogoliubov coefficients for both spin-3/2 and spin-1/2 will be
the same. Therefore, the general behavior of the generated
entanglement will also be the same.

Investigation of the generated entanglement in an expanding universe
for spin-0 and spin-1/2 particles shows some deep differences
between these modes. Our consideration in this paper proposes that
these differences are independent from the absolute value of the
spin of the particles. We argue that the absolute value of a spin
does not play any role in entanglement generation and the
differences are due to the bosonic or fermionic properties.



\begin{thebibliography}{99}
\bibitem{peres}
A. Peres, and D. R. Terno, Rev. Mod. Phys. {\bf 76}, 93 (2004).

\bibitem{alsing1}
 P. M. Alsing, and G. J. Milburn, Phys. Rev. Lett. {\bf 91}, 180404 (2003).

\bibitem{shi}
Y. Shi, Phys. Rev. D {\bf 70}, 105001 (2004).

\bibitem{alsing2}
P. M. Alsing, I. Fuentes-Schuller, R. B. Mann, and T. E. Tessier,
Phys. Rev. A {\bf 74}, 032326 (2006).

\bibitem{He}
S. He, S. Shao, and H. Zhang, J. Phys. A: Math. Theor. {\bf 40}, 857
(2007).

\bibitem{Aspect}
A. Aspect, J. Dalibard and G. Roger, Phys. Rev. Lett. {\bf 49}, 1804(1982).

\bibitem{Weihs}
G. Weihs, T. Jennewein, C. Simon, H. Weinfurter, and A. Zeilinger,
Phys. Rev. Lett. {\bf 81}, 5039 (1998).

\bibitem{Bouwmeester}
D. Bouwmeester, J. W. Pan, K. Mattle, M. Eibl, H. Weinfurter, and A.
Zeilinger, Nature {\bf 390}, 575 (1997).

\bibitem{Furusawa}
A. Furusawa, J. L. Sorensen, S. L. Braunstein, C. A. Fuchs, H. J.
Kimble, and E. S. Polzik, Science {\bf 282}, 706 (1998).

\bibitem{Fuentes1}
I. Fuentes-Schuller, and R. B. Mann, Phys. Rev. Lett. {\bf 95},
120404 (2005).

\bibitem{Fuentes2}
P. M. Alsing, I. Fuentes-Schuller, R. B. Mann, and T. E. Tessier,
Phys. Rev. A {\bf 74}, 032326 (2006).

\bibitem{mehri}
H. Mehri-Dehnavi, B. Mirza, H. Mohammadzadeh, and R. Rahimi, Ann.
Phys. {\bf 326}, 1320 (2011).

\bibitem{footnote}
For a density matrix $\rho$ of a composite bipartite
system $AB$, a separable state can be written as follows
   $    \rho_{\rm sep}=\sum_iw_i \rho_A^i\otimes \rho_B^i,
    $ where, $w_i$'s are positive weights, and $\rho_A^i$'s and
$\rho_B^i$'s are local states belonging to $A$ and $B$,
respectively. An entangled state  is a state that is not
separable.

\bibitem{Fuentes3}
J. L. Ball, I. Fuentes-Schuller, and F. P. Schuller, Phys. Lett. A
{\bf 359}, 550 (2006).

\bibitem{Fuentes4}
I. Fuentes, R. B. Mann, E. Martin-Martinez, and S. Moradi, Phys.
Rev. D. {\bf 82}, 045030 (2010).

\bibitem{kemmer}
N. Kemmer, Proc. Roy. Soc. A {\bf 177}, 9 (1939).

\bibitem{weinberg}
S. Weinberg, Phys. Rev. {\bf 133}, B1318 (1964).

\bibitem{shay}
S. Shay, and R. H. Good, Jr., Phys. Rev. {\bf 179}, 1410 (1969).

\bibitem{farhad}
F. Zamani, and A. Mostafazadeh, J. Math. Phys. {\bf 50}, 052302
(2009).

\bibitem{mirza}
B. Mirza, R. Narimani, and M. Zare, Eur. Phys. J. C {\bf 48}, 641
(2006).

\bibitem{swansson}
J. A. Swansson, and B. H. J. McKellar, J. Phys. A: Math Gen. {\bf
34}, 1051 (2001).

\bibitem{falek}
M. Falek, and M. Mera, Commun. Theor. Phys. {\bf 50}, 587 (2008).

\bibitem{greiner}
W. Greiner, {\it Relativistic Quantum Mechanics}, Springer-Verlag, (1990).

\bibitem{falek1}
M. Falek, and M. Merad, AIP Conf. Proc. {\bf 1444}, 367 (2012).

\bibitem{handbook}
M. Abramowitz, and I. A. Stegun, {\it Handbook of Mathematical
Functions: with Formulas, Graphs, and Mathematical Tables}, Dover,
(1965).

\bibitem{Davies}
N. D. Birrell, and P. C. W. Davies, {\it Quantum Fields in Curved
Space}, CUP (1982).

\bibitem{rarita}
W. Rarita, and J. Schwinger, Phys. Rev. {\bf 60}, 61 (1941).

\bibitem{velo}
G. Velo, and D. Zwanziger, Phys. Rev. {\bf 186}, 1337 (1969).

\bibitem{maroto}
A. L. Maroto, and A. Mazumdar, Phys. Rev. Lett. {\bf 84}, 1655
(2000).

\bibitem{Duncan}
A. Duncan, Phys. Rev. D {\bf 17}, 964 (1978).





\end{thebibliography}
\end{document}